\documentclass[onecollarge,final,amsfonts,amsmath,amssymb,aps,indentfirst,latexsym,mathrsfs,preprint,preprintnumbers,showpacks]{svjour2}

\usepackage{amsfonts,amsmath,amssymb,indentfirst,latexsym,mathrsfs}
\usepackage[mathscr]{eucal}
\usepackage[english]{babel}
\usepackage[T1]{fontenc}
\usepackage[cp1250]{inputenc}

\def\beq{\begin{equation}}
\def\eeq{\end{equation}}
\def\beqs{\begin{equation*}}
\def\eeqs{\end{equation*}}
\def\beqa{\begin{eqnarray}}
\def\eeqa{\end{eqnarray}}
\def\beqas{\begin{eqnarray*}}
\def\eeqas{\end{eqnarray*}}

\begin{document}

\title{Dirac-Bergmann Procedure Having Regard\\ to Interaction for Light-Front Yukawa Model} 

\author{Jan {\.Z}ochowski}
\institute{Faculty of Physics, University of Bia{\l}ystok, ul. Cio{\l}kowskiego 1L, 15-245 Bia{\l}ystok, Poland\\
\email{j.zochowski@uwb.edu.pl}}

\maketitle


\begin{abstract}

In this work we applied the Dirac-Bergmann procedure to establish the Dirac brackets, which have regard to interaction for the light-front Yukawa model in $D=1+3$ dimensions. We made use of a simple matrix equation leading to solution of the set task, wherein the main problem was to calculate the inverse matrix to the array composed of the constraints for enabled interaction. Proposed device comes down to the usage of certain, rather elementary matrix series. Obtained result - Dirac brackets including interaction for the Yukawa model - embraces the interacting contributions with first and second powers of the fermionic - scalar coupling constant. It is interesting from the physical point of view for discussion on the structure and the properties of the (anti-) commutators of the interacting theories on the light-front hyper-surface after quantization. We compared obtained results to the computations coming from modified method of the quantization, inferred from the Heisenberg equations. The open problem is whether the inverse matrix to this one, generated by the constraints, gives the complete and exact or only approximate solution of the studied problem.\\

\end{abstract}

\section{Quantization Methods of Yukawa Model}
\label{sec1}
 
In this work we study the light-front Yukawa model with one fermionic and one scalar field \cite{b12}. It is described in $D=1+3$ dimensions by the Lagrangian density 
\beq
\label{ldym}
{\cal L}=
{\overline\Psi}\left(i\!\!\not\!\partial-M\right)\Psi-
g\phi{\overline\Psi}\Psi+
{1\over 2}\left(\partial\phi\right)^{2}-
{1\over 2}m^{2}\phi^{2}-
\lambda\phi^{4}. 
\eeq
Herein, the dimensionless parameter $g$ plays the role of the coupling constant between scalar field $\phi$ with mass $m$ and two fermionic ones: incoming $\Psi$ and outgoing ${\overline\Psi}$, both  with common mass $M$. The coupling constant for self-interaction of the scalar field is denoted as $\lambda$, likewise dimensionless. The expanded, on the grounds of the fermionic bispinor indexes, light-front Yukawa model Lagrangian density satisfies
\beq
\label{ldymlf}
{\cal L}=
i\sqrt{2}\;\Psi^{\dagger}_{+}\partial_{+}\Psi_{+}+
i\sqrt{2}\;\Psi^{\dagger}_{-}\partial_{-}\Psi_{-}+
{i\over{\sqrt{2}}}
\Psi^{\dagger}_{-}\gamma^{+}\gamma^{j}\partial_{j}\Psi_{+}+
{i\over{\sqrt{2}}}
\Psi^{\dagger}_{+}\gamma^{-}\gamma^{j}\partial_{j}\Psi_{-}-
\eeq
\beqs
-{1\over{\sqrt{2}}}M
\Psi^{\dagger}_{-}\gamma^{+}\Psi_{+}-
{1\over{\sqrt{2}}}M
\Psi^{\dagger}_{+}\gamma^{-}\Psi_{-}-
{1\over{\sqrt{2}}}g
\phi\;\Psi^{\dagger}_{-}\gamma^{+}\Psi_{+}-
{1\over{\sqrt{2}}}g
\phi\;\Psi^{\dagger}_{+}\gamma^{-}\Psi_{-}+
\eeqs
\beqs
+\left(\partial_{+}\phi\right)\left(\partial_{-}\phi\right)-
{1\over 2}\left(\partial_{j}\phi\right)\left(\partial_{j}\phi\right)-
{1\over 2}m^{2}\phi^{2}-
\lambda\phi^{4}.
\eeqs
The equations of motion for discussed model may be inferred from the principle of minimal action, which leads to 
\beq
\label{empsim}
2i\partial_{\pm}\Psi_{\pm}+
i\gamma^{\mp}\!\!\not\!\partial_{\perp}\Psi_{\mp}-
\left(M+g\phi\right)\gamma^{\mp}\Psi_{\mp}=0,
\eeq
\beq
\label{emphi}
\left(2\partial_{+}\partial_{-}-\Delta_{\perp}\right)\phi+
m^{2}\phi+4\lambda\phi^{3}+
{1\over{\sqrt{2}}}g
\left(\Psi^{\dagger}_{+}\gamma^{-}\Psi_{-}+
\Psi^{\dagger}_{-}\gamma^{+}\Psi_{+}\right)=0.
\eeq
The canonical momenta for the Yukawa model: 
\beq
\label{defcanmom}
\pi_{\alpha}=
{{\partial{\cal L}}\over{\partial(\partial_{+}\varphi^{\dagger}_{\alpha})}}, \;\;\;\;\;\; 
\pi^{\dagger}_{\alpha}=
{{\partial{\cal L}}\over{\partial(\partial_{+}\varphi_{\alpha})}},
\eeq 
where $\varphi_{\alpha}=\Psi_{\alpha}$ or $\psi_{\alpha}=\phi$ both denote fermionic and scalar fields, may be obtained from the Lorentz invariant Lagrangian density (\ref{ldymlf}). The index $\alpha=+,-$ refers to bispinors. This well-known method gives:
\beq
\label{cmphipsippsim}
\pi_{+}=0, \;\;\;\;\;\;
\pi^{\dagger}_{+}=
-i\sqrt{2}\;\Psi^{\dagger}_{+}, \;\;\;\;\;\;
\pi_{-}=0, \;\;\;\;\;\; 
\pi^{\dagger}_{-}=0, \;\;\;\;\;\;
\pi_{\phi}=\partial_{-}\phi.
\eeq  
For that reason, the light-front canonical formalism for the Yukawa model postulates the Poisson brackets, which, after the quantization, are converted to the below (anti-) commutators: 
\beq
\label{canquant1}
\left\{
\Psi_{\pm}\left(x^{+},{\bar x}\right) \; , \; \pi^{\dagger}_{\pm}\left(x^{+},{\bar y}\right)
\right\}=
i\delta^{(3)}\left({\bar x}-{\bar y}\right)\Lambda_{\pm},
\eeq
\beq
\label{canquant2}
\left\{
\Psi^{\dagger}_{\pm}\left(x^{+},{\bar x}\right) \; , \; \pi_{\pm}\left(x^{+},{\bar y}\right)
\right\}=
i\delta^{(3)}\left({\bar x}-{\bar y}\right)\Lambda_{\pm},
\eeq
\beq
\label{canquant3}
\left[
\phi\left(x^{+},{\bar x}\right) \; , \; \pi_{\phi}\left(x^{+},{\bar y}\right)
\right]=
i\delta^{(3)}\left({\bar x}-{\bar y}\right).
\eeq
These (anti-) commutators are solely non-vanishing in this case. But this shortly summarized canonical quantization method is not the only one \cite{b9,b10,b11}.

We propose \cite{b15} the light-front quantization method, deduced from the Heisenberg equations. These equations are determined by the kinematic operator of the translation $P^{+}$. This version of the quantization neglects the classical canonical relationships of the studied model. It commences straightforwardly from the quantum Heisenberg equations \cite{b13}. The Lagrangian density (\ref{ldymlf}) allows us to derive following component of the canonical energy-momentum tensor
\beq
\label{emtpp}
T^{++}=
i\sqrt{2}\Psi^{\dagger}_{+}\partial_{-}\Psi_{+}+
\left(\partial_{-}\phi\right)^{2}.
\eeq
The $T^{+-}$ one may be obtained in the same manner
\beq
\label{emtpm}
T^{+-}=
i\sqrt{2}\;\Psi^{\dagger}_{+}\partial_{+}\Psi_{+}+
{1\over 2}
\left(\partial_{j}\phi\right)\left(\partial_{j}\phi\right)+
{1\over 2}m^{2}\phi^{2}+
\lambda\phi^{4}.
\eeq
The generators $P^{\mu}$ for the directions $x^{\mu}$ are defined as 
\beq
\label{gentralfmu}
P^{\mu}\left(x^{+}\right)=\!
\int_{R^{3}}\!\!{d^{3}}{\bar y}\;
T^{+\mu}\left(x^{+},{\bar y}\right), \;\;\;\;\;\; 
\mu=+,-,j, \;\;\;\;\;\; 
j=1,2. 
\eeq 
Therefore, the Heisenberg equations for the light-front direction $x^{+}$ satisfy: 
\beq
\label{hempsim}
i\partial_{-}\Psi_{\pm}\left(x^{+},{\bar x}\right)=
\left[\Psi_{\pm}\left(x^{+},{\bar x}\right) \; , \; P_{-}\left(x^{+}\right)\right], \;\;\;\;\;\;  
i\partial_{-}\phi\left(x^{+},{\bar x}\right)=
\left[\phi\left(x^{+},{\bar x}\right) \; , \; P_{-}\left(x^{+}\right)\right]. 
\eeq 
Now, with the help of the relations (\ref{hempsim}), we may easy calculate the light-front hyper-surface $x^{+}=0$ commutators and anti-commutators for the Yukawa model:   
\beq
\label{acpsippsidp}
\left\{\Psi_{+}\left(x^{+},{\bar x}\right) \; , \; \Psi^{\dagger}_{+}\left(x^{+},{\bar y}\right)\right\}=
-{1\over{\sqrt{2}}}
\delta^{(3)}\left({\bar x}-{\bar y}\right)\Lambda_{+},
\eeq
\beq
\label{acpsippsip}
\left\{\Psi_{+}\left(x^{+},{\bar x}\right) \; , \; \Psi_{+}\left(x^{+},{\bar y}\right)\right\}=0, \;\;\;\;\;\; 
\left[\phi\left(x^{+},{\bar x}\right) \; , \; \Psi_{+}\left(x^{+},{\bar y}\right)\right]=0,
\eeq 
\beq
\label{acpsimpsidp}
\left\{\Psi_{-}\left(x^{+},{\bar x}\right) \; , \; \Psi^{\dagger}_{+}\left(x^{+},{\bar y}\right)\right\}=
\eeq
\beqs
=-{i\over{2\sqrt{2}}}
\left\{\not\!\partial^{x}_{\perp}-
\left[M+g\phi\left(x^{+},{\bar y}\right)\right]\right\}\gamma^{+}
\partial^{-1_{x}}_{-}\delta\left(x^{-}-y^{-}\right)
\delta^{(2)}\left({\bf x}_{\perp}-{\bf y}_{\perp}\right),
\eeqs
\beq
\label{acpsimpsip}
\left\{\Psi_{-}\left(x^{+},{\bar x}\right) \; , \; \Psi_{+}\left(x^{+},{\bar y}\right)\right\}=0, \;\;\;\;\;\;  
\left[\phi\left(x^{+},{\bar x}\right) \; , \; \Psi_{-}\left(x^{+},{\bar y}\right)\right]=0,
\eeq
\beq
\label{cphipmphi}
\left[\phi\left(x^{+},{\bar x}\right) \; , \; \partial^{y}_{-}\phi\left(x^{+},{\bar y}\right)\right]=
{i\over 2}
\delta^{(3)}\left({\bar x}-{\bar y}\right),
\eeq
where $\partial_{-}^{-1_{x}}$ is merely the indefinite integration with respect to the coordinate $x^{-}$. Let's notice, two following anti-commutators $\{\Psi_{-}(x^{+},{\bar x}),\Psi^{\dagger}_{-}(x^{+},{\bar y})\}$ and $\{\Psi_{-}(x^{+},{\bar x}),\Psi_{-}(x^{+},{\bar y})\}$ are undetermined in the light of this method. Next considerations included in \cite{b15} show, that the first of them reveals the singularity on the light-front hyper-surface.\\
 
\section{Dirac-Bergmann Procedure with Interaction}
\label{dbphri}

As it can be seen \cite{b1,b2,b3}, discussed model includes three primary constraints: 
\beq
\label{constraintsprimary}
\Phi_{1}=\pi_{-}^{\dagger}=0, \;\;\;\;\;\;
\Phi_{2}=\pi_{-}=0, \;\;\;\;\;\;
\Phi_{3}=\pi_{\phi}-\partial_{-}\phi=0.
\eeq
The principle of minimal action provided the Dirac equation of motion (\ref{empsim}). Therefore, the Yukawa model embraces two secondary constraints: 
\beq
\label{constraintssecondary4} 
\Phi_{4}=
2i\partial_{-}\Psi^{\dagger}_{-}-
i\Psi^{\dagger}_{+}\not\!\!{\stackrel{\leftarrow}{\partial}}_{\perp}\gamma^{-}+
\left(M+g\phi\right)\Psi^{\dagger}_{+}\gamma^{-}=0,
\eeq
\beq
\label{constraintssecondary5}
\Phi_{5}=
2i\partial_{-}\Psi_{-}+
i\gamma^{+}\not\!\partial_{\perp}\Psi_{+}-
\left(M+g\phi\right)\gamma^{+}\Psi_{+}=0, 
\eeq
where we put $\lambda=0$. From then, we discuss the model without the self-interaction of the scalar field. These equations permit to express the $\Psi_{-}$ and the $\Psi^{\dagger}_{-}$ fields by the dynamical degrees of freedom. The Dirac-Bergmann procedure \cite{b6,b7,b8} allows to pass through from the Poisson brackets, which may be incompatible with the set of the constraints, to the Dirac ones, consistent with these $\Phi_{j}$, where is $j=1,\dots,5$ in our case. Finally, for the light-front fields $\psi_{a}(x^{+},{\bar x})$, we have  
\beq
\label{dirbracketdef}
\left\{\psi_{\alpha}\left(x^{+},{\bar x}\right) \; , \; \psi_{\beta}\left(x^{+},{\bar y}\right)\right\}_{D}=
\left\{\psi_{\alpha}\left(x^{+},{\bar x}\right) \; , \; \psi_{\beta}\left(x^{+},{\bar y}\right)\right\}_{P}-
\eeq
\beqs
-\sum_{k,l=1}^{5}
\int{d^{3}{\bar w}}\int{d^{3}{\bar z}}\;
\left\{\psi_{\alpha}\left(x^{+},{\bar x}\right) \; , \; \Phi_{k}\left(x^{+},{\bar w}\right)\right\}_{P}
{\cal F}^{-1}_{kl}\left(x^{+},{\bar w},{\bar z}\right)
\left\{\Phi_{l}\left(x^{+},{\bar z}\right) \; , \; \psi_{\beta}\left(x^{+},{\bar y}\right)\right\}_{P}. 
\eeqs
The symbol ${\cal F}^{-1}(x^{+},{\bar w},{\bar z})$ makes the inverse matrix with elements defined as the Poisson brackets of the constraints, derived with help of the canonical quantization rules 
\beq
\label{fmatrix}
{\cal F}_{kl}\left(x^{+},{\bar w},{\bar z}\right)=
\left\{\Phi_{k}\left(x^{+},{\bar w}\right) \; , \; \Phi_{l}\left(x^{+},{\bar z}\right)\right\}_{P}.
\eeq 
The mentioned ${\cal F}(x^{+},{\bar w},{\bar z})$ may be obtained after some algebraic and analytic computations with the above expressions and the constraints: (\ref{constraintsprimary}), (\ref{constraintssecondary4}), (\ref{constraintssecondary5}). It leads to the matrix of the size $5\times{5}$, which obeys the pattern 
\beq
\label{fmatrixelements}
{\cal F}\left(x^{+},{\bar w},{\bar z}\right)=
\eeq
\beqs
=\left(\;
\begin{array}{ccccc}
0&\;\;0&\;\;0&\;\;0&\;\;2i\partial^{w}_{-}\Lambda_{-}\\
0&\;\;0&\;\;0&\;\;2i\partial^{w}_{-}\Lambda_{-}&\;\;0\\
0&\;\;0&\;\;-2\partial^{w}_{-}\Lambda_{-}&
\;\;-g\psi^{\dagger}_{+}\left(x^{+},{\bar z}\right)\Lambda_{-}&
\;\;g\psi_{+}\left(x^{+},{\bar z}\right)\Lambda_{-}\\
0&\;\;2i\partial^{w}_{-}\Lambda_{-}&
\;\;g\psi^{\dagger}_{+}\left(x^{+},{\bar w}\right)\Lambda_{-}&
\;\;0&\;\;{\xi}\left(x^{+},{\bar w},{\bar z}\right)\\ 
2i\partial^{w}_{-}\Lambda_{-}&\;\;0&
\;\;-g\psi_{+}\left(x^{+},{\bar w}\right)\Lambda_{-}&
\;\;-{\xi}\left(x^{+},{\bar w},{\bar z}\right)&\;\;0\\
\end{array}
\right)
\delta^{(3)}\left({\bar w}-{\bar z}\right),  
\eeqs
where ${\xi}(x^{+},{\bar w},{\bar z})$ is showed by 
\beq
\label{amatrixfunction}
{\xi}\left(x^{+},{\bar w},{\bar z}\right)=
-i\sqrt{2}\Delta^{w}_{\perp}\Lambda_{-}+
i\sqrt{2}\left[M+g\phi\left(x^{+},{\bar w}\right)\right]
\left[M+g\phi\left(x^{+},{\bar z}\right)\right]
\Lambda_{-} 
\eeq 
and the fermionic fields $\psi_{+}$ or $\psi_{+}^{\dagger}$ are the components of the bispinors $\Psi$ or $\Psi^{\dagger}$, respectively (see also
Appendix, moreover \cite{b4,b5}).  

Structure of the ${\cal F}(x^{+},{\bar w},{\bar z})$, which contains, among others, the differential operators, brings on some difficulties on the way of performing its inverse. The matrix ${\cal F}^{-1}(x^{+},{\bar x},{\bar y})$ may be readily and directly evaluated only in the case of the free fermionic and scalar fields. For the purpose of overcoming the mathematical troubles and obtaining the complete form of the 
${\cal F}^{-1}(x^{+},{\bar x},{\bar y})$ with $g\neq{0}$, we must decompose the studied matrix as the sum of two contributions
\beq
\label{fmatrixdecomp}
{\cal F}\left(x^{+},{\bar w},{\bar z}\right)=
{\cal F}_{0}\left(x^{+},{\bar w},{\bar z}\right)+
{\cal G}\left(x^{+},{\bar w},{\bar z}\right),
\eeq
where the first one, ${\cal F}_{0}(x^{+},{\bar w},{\bar z})$, includes only the free fields ($g=0$). The whole dependence on the interaction in this model is moved to the second matrix ${\cal G}(x^{+},{\bar w},{\bar z})$. Accordingly:
\beq
\label{fmatrixfreeelements}
{\cal F}_{0}\left(x^{+},{\bar w},{\bar z}\right)=
\eeq
\beqs
=\left(\;
\begin{array}{ccccc}
0&\;\;0&\;\;0&\;\;0&\;\;2i\partial^{w}_{-}\Lambda_{-}\\
0&\;\;0&\;\;0&\;\;2i\partial^{w}_{-}\Lambda_{-}&\;\;0\\
0&\;\;0&\;\;-2\partial^{w}_{-}\Lambda_{-}&\;\;0&\;\;0\\
0&\;\;2i\partial^{w}_{-}\Lambda_{-}&\;\;\;0&\;\;0&
\;\;{\xi}_{0}\left(x^{+},{\bar w},{\bar z}\right)\\ 
2i\partial^{w}_{-}\Lambda_{-}&\;\;0&\;\;\;0&
\;\;-{\xi}_{0}\left(x^{+},{\bar w},{\bar z}\right)&\;\;0\\
\end{array}
\right)
\delta^{(3)}\left({\bar w}-{\bar z}\right), 
\eeqs  
wherein the operator ${\xi}_{0}(x^{+},{\bar w},{\bar z})$ descends from those ${\xi}(x^{+},{\bar w},{\bar z})$, in which the coupling constant was put as equal to zero 
\beq
\label{amatrixfunctionzero}
{\xi}_{0}\left(x^{+},{\bar w},{\bar z}\right)=
-i\sqrt{2}\Delta^{w}_{\perp}\Lambda_{-}+
i\sqrt{2}M^{2}\Lambda_{-}. 
\eeq
The complete information about the interactions in this model is inserted into the matrix ${\cal G}(x^{+},{\bar w},{\bar z})$, which has the form
\beq
\label{gmatrixelements}
{\cal G}\left(x^{+},{\bar w},{\bar z}\right)=
\eeq
\beqs
=\left(\;
\begin{array}{ccccc}
0&\;\;0&\;\;0&\;\;0&\;\;0\\
0&\;\;0&\;\;0&\;\;0&\;\;0\\
0&\;\;0&\;\;0&\;\;-g\psi^{\dagger}_{+}\left(x^{+},{\bar z}\right)\Lambda_{-}&
\;\;g\psi_{+}\left(x^{+},{\bar z}\right)\Lambda_{-}\\
0&\;\;0&\;\;g\psi^{\dagger}_{+}\left(x^{+},{\bar w}\right)\Lambda_{-}&
\;\;\;0&\;\;{\xi}_{g}\left(x^{+},{\bar w},{\bar z}\right)\\ 
0&\;\;0&\;\;-g\psi_{+}\left(x^{+},{\bar w}\right)\Lambda_{-}&
\;\;-{\xi}_{g}\left(x^{+},{\bar w},{\bar z}\right)&\;\;0\\
\end{array}
\right)
\delta^{(3)}\left({\bar w}-{\bar z}\right). 
\eeqs
The element ${\xi}_{g}(x^{+},{\bar w},{\bar z})={\xi}(x^{+},{\bar w},{\bar z})-{\xi}_{0}(x^{+},{\bar w},{\bar z})$ is defined as an operator which embraces the interacting scalar field    
\beq
\label{amatrixfunctioninteract}
{\xi}_{g}\left(x^{+},{\bar w},{\bar z}\right)=
i\sqrt{2}gM
\left[
{\phi}\left(x^{+},{\bar w}\right)+{\phi}\left(x^{+},{\bar z}\right)
\right]
\Lambda_{-}+
i\sqrt{2}g^{2}
{\phi}\left(x^{+},{\bar w}\right){\phi}\left(x^{+},{\bar z}\right)\Lambda_{-}. 
\eeq

Among many exploitable algebraic formulas, we have particularly interesting, which express the reverse sum of the matrices by the series of their inverses and themselves   
\beq
\label{matrixseries}
\left(A+S\right)^{-1}=
A^{-1}+
\sum_{n=1}^{+\infty}
(-1)^{n}\left(A^{-1}S\right)^{n}\!A^{-1}.
\eeq
The mathematical requirements for the correctness of this equation embrace invertibility of the matrix $A+S$ and $A$. The matrix $S$ does not have to posses the $S^{-1}$. It may be even singular, as in the case of our considerations. The above equation may provide finite case if the right side series was perfectly truncated, due to zeroth product of the relevant matrices. Therefore, it gives exact solution for the reverse sum of the left side of discussed formula (\ref{matrixseries}). But even for the infinite version of this pattern, we can handle the matrix $S$ as comprising small parameter, the dimensionless coupling constant $g\ll{1}$, for instance. It also allows to do effective calculations, physically treated as the perturbations. Discussed ${\cal F}(x^{+},{\bar w},{\bar z})$ embraces the differential operators acting onto the Dirac deltas, ergo its inverse ought to be defined with presence of the integrals over convoluted coordinates on the light-front hyper-surface  
\beq
\label{matrixintegralinversion} 
\int_{R^{3}}\!{d^{3}{\bar z}}\;
{\cal F}^{-1}\!\left(x^{+},{\bar x},{\bar z}\right)
{\cal F}\left(x^{+},{\bar z},{\bar y}\right)=
\delta^{(3)}\left({\bar x}-{\bar y}\right)I=
\int_{R^{3}}\!{d^{3}{\bar z}}\;
{\cal F}\left(x^{+},{\bar x},{\bar z}\right)
{\cal F}^{-1}\!\left(x^{+},{\bar z},{\bar y}\right). 
\eeq
Symbol $I$ denotes the unit array of the relevant size. If we are able to obtain the exact result for the inverse free matrix 
${\cal F}_{0}^{-1}(x^{+},{\bar x},{\bar y})$, then the complete expression for the ${\cal F}^{-1}(x^{+},{\bar x},{\bar y})$ shall be displayed with help of the pattern (\ref{matrixseries}), as the following chain of the matrices, where each product of them is affiliated to the integration 
\beq
\label{matrixseries}
{\cal F}^{-1}\!\left(x^{+},{\bar x},{\bar y}\right)=
\left[
{\cal F}_{0}\left(x^{+},{\bar x},{\bar y}\right)+
{\cal G}\left(x^{+},{\bar x},{\bar y}\right)\right]^{-1}=
{\cal F}_{0}^{-1}\!\left(x^{+},{\bar x},{\bar y}\right)+
\eeq
\beqs
+\!\sum_{n=1}^{+\infty}
(-1)^{n}\!\!
\int_{R^{3}}\!\!\!\!{d^{3}{{\bar z}_{1}}}
\dots\!\!
\int_{R^{3}}\!\!\!\!{d^{3}{{\bar z}_{n+1}}}
\left\{
{\cal F}_{0}^{-1}\!\left(x^{+},{\bar x},{\bar z}_{1}\right)
{\cal G}\left(x^{+},{\bar z}_{1},{\bar z}_{2}\right)\dots
{\cal F}_{0}^{-1}\!\left(x^{+},{\bar z}_{j-1},{\bar z}_{j}\right)
{\cal G}\left(x^{+},{\bar z}_{j},{\bar z}_{j+1}\right)\dots
\right.
\eeqs
\beqs
\left.
\dots{\cal F}_{0}^{-1}\!\left(x^{+},{\bar z}_{n-1},{\bar z}_{n}\right)
{\cal G}\left(x^{+},{\bar z}_{n},{\bar z}_{n+1}\right)
\right\}
{\cal F}_{0}^{-1}\!\left(x^{+},{\bar z}_{n+1},{\bar y}\right).
\eeqs
The matrix ${\cal G}(x^{+},{\bar w},{\bar z})$ includes the coupling constant to the power of one and two. On account of that we rewrite this ${\cal G}(x^{+},{\bar w},{\bar z})$ as the sum of two matrices terms proportional to the $g$ and $g^{2}$, what enables us to do the computations with desired clarity 
\beq
\label{matrixg12}
{\cal G}\left(x^{+},{\bar w},{\bar z}\right)=
g\;\!{\cal{\tilde G}}_{1}\left(x^{+},{\bar w},{\bar z}\right)+
g^{2}\;\!{\cal{\tilde G}}_{2}\left(x^{+},{\bar w},{\bar z}\right).
\eeq
In this way, we have introduced two new matrices ${\cal{\tilde G}}_{1}(x^{+},{\bar w},{\bar z})$ and 
${\cal{\tilde G}}_{2}(x^{+},{\bar w},{\bar z})$. Structure of the first of them is easily displayed below 
\beq
\label{gtildematrixelement1}
{\cal{\tilde G}}_{1}\left(x^{+},{\bar w},{\bar z}\right)=
\eeq
\beqs
=\left(\;
\begin{array}{ccccc}
0&\;\;0&\;\;0&\;\;0&\;\;0\\
0&\;\;0&\;\;0&\;\;0&\;\;0\\
0&\;\;0&\;\;0&\;\;-\psi^{\dagger}_{+}\left(x^{+},{\bar z}\right)\Lambda_{-}&
\;\;\psi_{+}\left(x^{+},{\bar z}\right)\Lambda_{-}\\
0&\;\;0&\;\;\psi^{\dagger}_{+}\left(x^{+},{\bar w}\right)\Lambda_{-}&
\;\;\;0&\;\;{\xi}_{1}\left(x^{+},{\bar w},{\bar z}\right)\\ 
0&\;\;0&\;\;-\psi_{+}\left(x^{+},{\bar w}\right)\Lambda_{-}&
\;\;-{\xi}_{1}\left(x^{+},{\bar w},{\bar z}\right)&\;\;0\\
\end{array}
\right)
\delta^{(3)}\left({\bar w}-{\bar z}\right),
\eeqs
where another ${\xi}_{1}(x^{+},{\bar w},{\bar z})$ comes from already defined ${\xi}_{g}(x^{+},{\bar w},{\bar z})$ and yields
\beq
\label{ksi1}
{\xi}_{1}\left(x^{+},{\bar w},{\bar z}\right)=
i\sqrt{2}M
\left[
{\phi}\left(x^{+},{\bar w}\right)+
{\phi}\left(x^{+},{\bar z}\right)
\right]\Lambda_{-}. 
\eeq
The other, matrix ${\cal{\tilde G}}_{2}(x^{+},{\bar w},{\bar z})$, obeys the pattern 
\beq
\label{gtildematrixelement2}
{\cal{\tilde G}}_{2}\left(x^{+},{\bar w},{\bar z}\right)=
\eeq
\beqs
=\left(\;
\begin{array}{ccccc}
0&\;\;0&\;\;0&\;\;0&\;\;0\\
0&\;\;0&\;\;0&\;\;0&\;\;0\\
0&\;\;0&\;\;0&\;\;0&\;\;0\\
0&\;\;0&\;\;0&\;\;0&\;\;{\xi}_{2}\left(x^{+},{\bar w},{\bar z}\right)\\ 
0&\;\;0&\;\;0&\;\;-{\xi}_{2}\left(x^{+},{\bar w},{\bar z}\right)&\;\;0\\
\end{array}
\right)
\delta^{(3)}\left({\bar w}-{\bar z}\right),
\eeqs   
with ${\xi}_{2}(x^{+},{\bar w},{\bar z})$ being equal to 
${\xi}_{2}(x^{+},{\bar w},{\bar z})={\xi}_{g}(x^{+},{\bar w},{\bar z})-{\xi}_{1}(x^{+},{\bar w},{\bar z}) $, what finally gives  
\beq
\label{ksi2}
{\xi}_{2}\left(x^{+},{\bar w},{\bar z}\right)=
i\sqrt{2}
{\phi}\left(x^{+},{\bar w}\right)
{\phi}\left(x^{+},{\bar z}\right)
\Lambda_{-}. 
\eeq 
The general prescription (\ref{matrixseries}) for effective obtaining the inverse matrix ${\cal F}^{-1}(x^{+},{\bar x},{\bar y})$ should be, with the assistance of the presented decomposition of the ${\cal G}(x^{+},{\bar w},{\bar z})$ onto two terms (\ref{matrixg12}), reformulated as the series with sequent powers of the $g$. We insert the equation (\ref{matrixg12}) into the said general expansion (\ref{matrixseries}) and therefore, we achieve at the end the series  
\beq
\label{finverseperturbative} 
{\cal F}^{-1}\left(x^{+},{\bar x},{\bar y}\right)=
{\cal F}_{0}^{-1}\left(x^{+},{\bar x},{\bar y}\right)+
\sum_{n=1}^{+\infty}
{\cal B}_{n}\left(x^{+},{\bar x},{\bar y}\right).  
\eeq
The successive orders of presented calculations satisfy:  
\beq
\label{finverseperturbativen}
{\cal B}_{n}\left(x^{+},{\bar x},{\bar y}\right)=
\eeq
\beqs
=\!g^{n}\!\!\!
\sum_{j_{n}=1}^{n}\!\!
(-1)^{j_{n}}\!\!\;\!
\left\{\!
{\prod_{k=1}^{j_{n}}}{{ }^{{ }^{\prime}}}\!\!
\left[
\int_{R^{3}}\!\!{d^{3}{\bar w}_{k}}\!\!
\int_{R^{3}}\!\!{d^{3}{\bar z}_{k}}\;\!
{\cal F}_{0}^{-1}\!\left(x^{+},{\bar w}_{k-1},{\bar z}_{k}\right)
{\cal{\tilde G}}_{{\bar\nu}_{2}^{k}}\!\left(x^{+},{\bar z}_{k},{\bar w}_{k+1}\right)
\right]_{{{\bar w}_{0}={\bar x}}}\!
\right\}
{\cal F}_{0}^{-1}\!\left(x^{+},{\bar w}_{j_{n}+1},{\bar y}\right)\!.
\eeqs
The prime introduced at the symbol of the product means, that the sum of the indexes in the chain of the matrices ${\cal{\tilde G}}_{1,2}$ should be equal to $n$, which determines order of the expansion. Therefore, mentioned prime denotes the request ${\bar\nu}_{2}^{j_{1}}+\dots{+}{\bar\nu}_{2}^{j_{n}}=n$ for the $j_{l}$ - the tuples of the two-element set, being the ${\cal{\tilde G}}$ indexes. Herein $l=1,\dots{,}n$. The Dirac-Bergmann procedure, described above, allows to introduce the interaction into the canonical formalism of the studied model.\\  

\section{Inverse Free Matrix ${\cal F}_{0}^{-1}$}
\label{ifmf}

We should commence the Dirac-Bergmann procedure from the calculation of the free inverse matrix ${\cal F}_{0}^{-1}(x^{+},{\bar x},{\bar y})$. Observation, that the structure of the matrix (\ref{fmatrixfreeelements}) is the anti-down-triangle, allows us to postulate, according to relevant algebraic theorem, that its inverse has the form of the anti-up-triangle array 
\beq
\label{invfmatrixelements} 
{\cal F}_{0}^{-1}\left(x^{+},{\bar x},{\bar y}\right)=
{\cal F}_{0}^{-1}\left(x^{+},{\bar x}-{\bar y}\right)=
\eeq
\beqs
=\left(
\begin{array}{ccccc}
0&\;\;c\left(x^{+},{\bar x}-{\bar y}\right)I&\;\;0&\;\;0&
\;\;a\left(x^{+},{\bar x}-{\bar y}\right)I\\
-c\left(x^{+},{\bar x}-{\bar y}\right)I&\;\;0&\;\;0&
\;\;a\left(x^{+},{\bar x}-{\bar y}\right)I&\;\;0\\
0&\;\;0&\;\;b\left(x^{+},{\bar x}-{\bar y}\right)I&\;\;0&\;\;0\\
0&\;\;a\left(x^{+},{\bar x}-{\bar y}\right)I&\;\;0&\;\;0&\;\;0\\ 
a\left(x^{+},{\bar x}-{\bar y}\right)I&\;\;0&\;\;0&\;\;0&\;\;0\\
\end{array}
\right),  
\eeqs
with $I$ denoting the unit matrix of the size $2\times{2}$. The above ${\cal F}_{0}^{-1}(x^{+},{\bar x},{\bar y})$ describes the case without interactions and on account of that, it has the explicit translational symmetry. The existence of the inverse matrix
(\ref{matrixintegralinversion}) is realized under the conditions, that the functions $a(x^{+},{\bar x})$, $b(x^{+},{\bar x})$ and 
$c(x^{+},{\bar x})$ obey the differential equations: 
\beq
\label{abdiffeq}
\partial_{-}a\left(x^{+},{\bar x}\right)=
-{i\over 2}\delta^{(3)}\left({\bar x}\right), \;\;\;\;\;\;  
\partial_{-}b\left(x^{+},{\bar x}\right)=
-{1\over 2}\delta^{(3)}\left({\bar x}\right), \;\;\;\;\;\;  
\eeq
\beq
\label{cdiffeq}
\partial_{-}c\left(x^{+},{\bar x}\right)+
{1\over{\sqrt{2}}}
\left(\Delta_{\perp}-M^{2}\right)
a\left(x^{+},{\bar x}\right)=0
\eeq
These functions are not set down uniquely. Any boundary conditions, imposed on the functions $a(x^{+},{\bar x})$ and $c(x^{+},{\bar x})$ and doing them unambiguous, permit also in this way to establish the well-defined Dirac brackets onto the light-front hyper-surface. But mentioned boundary conditions seem to be not clear from the physical point of view. They may bring on violation of the Lorentz or eventually internal symmetry of the studied theory \cite{b14}. It is possible to obtain the explicit solution for $a(x^{+},{\bar x})$ and $b(x^{+},{\bar x})$ only in the simplest case. The anti-symmetry of the relevant Dirac brackets allows us to write, that $a(x^{+},{\bar x})=-(i/4){\rm sgn}(x^{-})\delta^{(2)}({\bf x}_{\perp})$ and analogously $b(x^{+},{\bar x})=-(1/4){\rm sgn}(x^{-})\delta^{(2)}({\bf x}_{\perp})$, what is consistent with the result (\ref{acpsimpsidp}).

Using the prescription (\ref{dirbracketdef}), we can unequivocally compute the following nontrivial Dirac brackets for the light-front Yukawa model with the interaction switched off ($g=0$). Taking all the Poisson brackets from the canonical light-front formalism for the Yukawa model, we put:  
\beq
\label{dirbracketpsippsidp}
\left\{
\Psi_{+}\left(x^{+},{\bar x}\right) \; , \; \Psi^{\dagger}_{+}\left(x^{+},{\bar y}\right)
\right\}_{D}=
\left\{
\Psi_{+}\left(x^{+},{\bar x}\right) \; , \; \Psi^{\dagger}_{+}\left(x^{+},{\bar y}\right)
\right\}_{P}=
{i\over{\sqrt{2}}}
\delta^{(3)}\left({\bar x}-{\bar y}\right)
\Lambda_{+},
\eeq
\beq
\label{dirbracketphiphi}
\left\{
\phi\left(x^{+},{\bar x}\right) \; , \; \phi\left(x^{+},{\bar y}\right)
\right\}_{D}=
{\underbrace{
\left\{
\phi\left(x^{+},{\bar x}\right) \; , \; \phi\left(x^{+},{\bar y}\right)
\right\}_{P}
}_{0}}+
b\left(x^{+},{\bar x}-{\bar y}\right), 
\eeq
\beq
\label{dirbracketphiparmphi}
\left\{
\phi\left(x^{+},{\bar x}\right) \; , \; \partial^{y}_{-}\phi\left(x^{+},{\bar y}\right)
\right\}_{D}=
{\underbrace{
\left\{
\phi\left(x^{+},{\bar x}\right) \; , \; \partial^{y}_{-}\phi\left(x^{+},{\bar y}\right)
\right\}_{P}
}_{0}}+
\partial_{-}^{y}b\left(x^{+},{\bar x}-{\bar y}\right)=
{1\over 2}
\delta^{(3)}\left({\bar x}-{\bar y}\right).
\eeq
The two below Dirac brackets depend on the functions $a(x^{+},{\bar x})$ and $c(x^{+},{\bar x})$. They are described by the formulas:
\beq
\label{dirbracketpsimpsidp}
\left\{
\Psi_{-}\left(x^{+},{\bar x}\right) \; , \; \Psi^{\dagger}_{+}\left(x^{+},{\bar y}\right)
\right\}_{D}=
{\underbrace{
\left\{
\Psi_{-}\left(x^{+},{\bar x}\right) \; , \; \Psi^{\dagger}_{+}\left(x^{+},{\bar y}\right)
\right\}_{P}
}_{0}}-
\eeq
\beqs
-{1\over{\sqrt{2}}}
\left[
i\!\not\!\partial^{x}_{\perp}a\left(x^{+},{\bar x}-{\bar y}\right)-
iMa\left(x^{+},{\bar x}-{\bar y}\right)
\right]
\gamma^{+},
\eeqs
\beq
\label{dirbracketpsimpsidm}
\left\{
\Psi_{-}\left(x^{+},{\bar x}\right) \; , \; \Psi^{\dagger}_{-}\left(x^{+},{\bar y}\right)
\right\}_{D}=
{\underbrace{
\left\{
\Psi_{-}\left(x^{+},{\bar x}\right) \; , \; \Psi^{\dagger}_{-}\left(x^{+},{\bar y}\right)
\right\}_{P}
}_{0}}+
c\left(x^{+},{\bar x}-{\bar y}\right)
\Lambda_{-}.
\eeq
The Dirac brackets of the other combinations of the fermionic fields are equal the relevant Poisson ones: 
\beq
\label{dirbracketpsippsip}
\left\{
\Psi_{+}\left(x^{+},{\bar x}\right) \; , \; \Psi_{+}\left(x^{+},{\bar y}\right)
\right\}_{D}=
\left\{
\Psi_{+}\left(x^{+},{\bar x}\right) \; , \; \Psi_{+}\left(x^{+},{\bar y}\right)
\right\}_{P}=0, 
\eeq
\beq
\label{dirbracketpsimpsip}
\left\{
\Psi_{+}\left(x^{+},{\bar x}\right) \; , \; \Psi_{-}\left(x^{+},{\bar y}\right)
\right\}_{D}=
\left\{
\Psi_{+}\left(x^{+},{\bar x}\right) \; , \; \Psi_{-}\left(x^{+},{\bar y}\right)
\right\}_{P}=0, 
\eeq
\beq
\label{dirbracketpsimpsim}
\left\{
\Psi_{-}\left(x^{+},{\bar x}\right) \; , \; \Psi_{-}\left(x^{+},{\bar y}\right)
\right\}_{D}=
\left\{
\Psi_{-}\left(x^{+},{\bar x}\right) \; , \; \Psi_{-}\left(x^{+},{\bar y}\right)
\right\}_{P}=0. 
\eeq
Also the mixed Dirac brackets for both the scalar and fermionic fields, after the application of the Dirac-Bergmann procedure, vanish: 
\beq
\label{dirbracketphipsip}
\left\{
\phi\left(x^{+},{\bar x}\right) \; , \; \Psi_{+}\left(x^{+},{\bar y}\right)
\right\}_{D}=
\left\{
\phi\left(x^{+},{\bar x}\right) \; , \; \Psi_{+}\left(x^{+},{\bar y}\right)
\right\}_{P}=0,  
\eeq
\beq
\label{dirbracketphipsim}
\left\{
\phi\left(x^{+},{\bar x}\right) \; , \; \Psi_{-}\left(x^{+},{\bar y}\right)
\right\}_{D}=
\left\{
\phi\left(x^{+},{\bar x}\right) \; , \; \Psi_{-}\left(x^{+},{\bar y}\right)
\right\}_{P}=0. 
\eeq
Our issues have nearly analogous form to those, derived by a modified method of the quantization and described in Chapter \ref{sec1}. Remarkably, the pattern (\ref{dirbracketpsimpsidp}) is consistent with the anti-commutator (\ref{acpsimpsidp}), taken for $g=0$. Significant change we observe for the expression (\ref{dirbracketpsimpsidm}), which differs about the term with function $c(x^{+},{\bar x})$  in comparison to standard results. But we must do at this point certain reservation, that modified quantization method \cite{b15}, based on the Heisenberg equations, leads to the undetermined anti-commutators $\{\Psi_{-}(x^{+},{\bar x}),\Psi^{\dagger}_{-}(x^{+},{\bar y})\}$ and 
$\{\Psi_{-}(x^{+},{\bar x}),\Psi_{-}(x^{+},{\bar y})\}$. Moreover, the next analysis in this work notes, that the anti-commutator $\{\Psi_{-}(x^{+},{\bar x}),\Psi^{\dagger}_{-}(x^{+},{\bar y})\}$ is singular on the light-front hyper-surface in a view of the modified quantization method.\\

\section{Inverse Matrix ${\cal F}^{-1}$ with Interaction}
\label{ifmi}

Let's employ the expression (\ref{finverseperturbative}) to perform detail study of the light-front Dirac brackets with interaction for the Yukawa model. We can write down three lowest orders of the expansion of the full matrix ${\cal F}^{-1}$ in the powers of coupling constant 
${\cal F}^{-1}={\cal F}_{0}^{-1}+{\cal B}_{1}+{\cal B}_{2}+{\cal B}_{3}$. It shall be clear afterward, that the chosen power of the expansion in this case is not accidental. After some algebraic computations we obtain: 
\beq
\label{expfmionr1}
{\cal B}_{1}\!\left(x^{+},{\bar x},{\bar y}\right)=
-g
\left\{
\int_{R^{3}}\!{d^{3}{\bar z}_{1}}\!
\int_{R^{3}}\!{d^{3}{\bar z}_{2}}\!\;
{\cal F}_{0}^{-1}\!\left({\bar x}-{\bar z}_{1}\right)
{\cal{\tilde G}}_{1}\!\left({\bar z}_{1},{\bar z}_{2}\right)
{\cal F}_{0}^{-1}\!\left({\bar z}_{2}-{\bar y}\right)
\right\},
\eeq
\beq
\label{expfmionr2}
{\cal B}_{2}\!\left(x^{+},{\bar x},{\bar y}\right)=
g^{2}
\left\{
-\int_{R^{3}}\!{d^{3}{\bar z}_{1}}\!
\int_{R^{3}}\!{d^{3}{\bar z}_{2}}\!\;
{\cal F}_{0}^{-1}\!\left({\bar x}-{\bar z}_{1}\right)
{\cal{\tilde G}}_{2}\!\left({\bar z}_{1},{\bar z}_{2}\right)
{\cal F}_{0}^{-1}\!\left({\bar z}_{2}-{\bar y}\right)+
\right.
\eeq
\beqs
\left.
+\int_{R^{3}}\!{d^{3}{\bar z}_{1}}\dots\!
\int_{R^{3}}\!{d^{3}{\bar z}_{4}}\!\;
{\cal F}_{0}^{-1}\left({\bar x}-{\bar z}_{1}\right)
{\cal{\tilde G}}_{1}\!\left({\bar z}_{1},{\bar z}_{2}\right)
{\cal F}_{0}^{-1}\!\left({\bar z}_{2}-{\bar z}_{3}\right)
{\cal{\tilde G}}_{1}\!\left({\bar z}_{3},{\bar z}_{4}\right)
{\cal F}_{0}^{-1}\!\left({\bar z}_{4}-{\bar y}\right)
\right\},
\eeqs
\beq
\label{expfmionr3}
{\cal B}_{3}\!\left(x^{+},{\bar x},{\bar y}\right)=
\eeq
\beqs
=g^{3}
\left\{
\int_{R^{3}}\!{d^{3}{\bar z}_{1}}\dots\!
\int_{R^{3}}\!\!{d^{3}{\bar z}_{4}}\!\;
{\cal F}_{0}^{-1}\!\left({\bar x}-{\bar z}_{1}\right)
{\cal{\tilde G}}_{1}\!\left({\bar z}_{1},{\bar z}_{2}\right)
{\cal F}_{0}^{-1}\!\left({\bar z}_{2}-{\bar z}_{3}\right)
{\cal{\tilde G}}_{2}\!\left({\bar z}_{3},{\bar z}_{4}\right)
{\cal F}_{0}^{-1}\!\left({\bar z}_{4}-{\bar y}\right)+
\right.
\eeqs
\beqs
+\int_{R^{3}}\!{d^{3}{\bar z}_{1}}\dots\!
\int_{R^{3}}\!{d^{3}{\bar z}_{4}}\!\;
{\cal F}_{0}^{-1}\!\left({\bar x}-{\bar z}_{1}\right)
{\cal{\tilde G}}_{2}\!\left({\bar z}_{1},{\bar z}_{2}\right)
{\cal F}_{0}^{-1}\!\left({\bar z}_{2}-{\bar z}_{3}\right)
{\cal{\tilde G}}_{1}\!\left({\bar z}_{3},{\bar z}_{4}\right)
{\cal F}_{0}^{-1}\!\left({\bar z}_{4}-{\bar y}\right)-
\eeqs
\beqs
\left.
-\!\!\int_{R^{3}}\!\!\!\!{d^{3}{\bar z}_{1}}\dots
\!\!\int_{R^{3}}\!\!\!\!{d^{3}{\bar z}_{6}}\!\;
{\cal F}_{0}^{-1}\!\left({\bar x}-{\bar z}_{1}\right)\!
{\cal{\tilde G}}_{1}\!\left({\bar z}_{1},{\bar z}_{2}\right)\!
{\cal F}_{0}^{-1}\!\left({\bar z}_{2}-{\bar z}_{3}\right)\!
{\cal{\tilde G}}_{1}\!\left({\bar z}_{3},{\bar z}_{4}\right)\!
{\cal F}_{0}^{-1}\!\left({\bar z}_{4}-{\bar z}_{5}\right)\!
{\cal{\tilde G}}_{1}\!\left({\bar z}_{5},{\bar z}_{6}\right)\!
{\cal F}_{0}^{-1}\!\left({\bar z}_{6}-{\bar y}\right)
\right\}\!.
\eeqs
It is easy to convince, that the first order prescription (\ref{expfmionr1}) gives, together with the array (\ref{gtildematrixelement1}) and the expression (\ref{ksi1}), following result
\beq
\label{invfinr1} 
{\cal B}_{1}\!\left(x^{+},{\bar x},{\bar y}\right)=
-g
\left(
\begin{array}{ccccc}
0&\;\;F_{1}\left(x^{+},{\bar x},{\bar y}\right)&
\;\;F_{2}\left(x^{+},{\bar x},{\bar y}\right)&\;\;0&\;\;0\\
-F_{1}\left(x^{+},{\bar x},{\bar y}\right)&\;\;0&\;\;
F_{3}\left(x^{+},{\bar x},{\bar y}\right)&\;\;0&\;\;0\\
F_{4}\left(x^{+},{\bar x},{\bar y}\right)&\;\;
F_{5}\left(x^{+},{\bar x},{\bar y}\right)&\;\;&\;\;0&\;\;0\\
0&\;\;0&\;\;0&\;\;0&\;\;0\\ 
0&\;\;0&\;\;0&\;\;0&\;\;0\\
\end{array}
\right).
\eeq
The elements inside this array are containing the components of the fermionic field and two already discussed functions: $a(x^{+},{\bar x})$, $b(x^{+},{\bar x})$, coming from the inverse ${\cal F}_{0}^{-1}(x^{+},{\bar x},{\bar y})$ matrix:    
\beq
\label{invfinr1elements1}
F_{1}\left(x^{+},{\bar x},{\bar y}\right)=
2\sqrt{2}iM\!\!
\int_{R^{3}}\!{d^{3}{\bar z}}\;
a\left(x^{+},{\bar x}-{\bar z}\right)
{\phi}\left(x^{+},{\bar z}\right)
a\left(x^{+},{\bar z}-{\bar y}\right)
\Lambda_{-},
\eeq
\beq
\label{invfinr1elements2}
F_{2}\left(x^{+},{\bar x},{\bar y}\right)=
-\!\int_{R^{3}}\!{d^{3}{\bar z}}\;
a\left(x^{+},{\bar x}-{\bar z}\right)
\psi_{+}\left(x^{+},{\bar z}\right)
b\left(x^{+},{\bar z}-{\bar y}\right)
\Lambda_{-},
\eeq 
\beq
\label{invfinr1elements3}
F_{3}\left(x^{+},{\bar x},{\bar y}\right)=
\int_{R^{3}}\!{d^{3}{\bar z}}\;
a\left(x^{+},{\bar x}-{\bar z}\right)
\psi_{+}^{\dagger}\left(x^{+},{\bar z}\right)
b\left(x^{+},{\bar z}-{\bar y}\right)
\Lambda_{-},
\eeq
\beq
\label{invfinr1elements4}
F_{4}\left(x^{+},{\bar x},{\bar y}\right)=
\int_{R^{3}}\!{d^{3}{\bar z}}\;
b\left(x^{+},{\bar x}-{\bar z}\right)
\psi_{+}\left(x^{+},{\bar z}\right)
a\left(x^{+},{\bar z}-{\bar y}\right)
\Lambda_{-},
\eeq
\beq
\label{invfinr1elements5}
F_{5}\left(x^{+},{\bar x},{\bar y}\right)=
-\!\int_{R^{3}}\!{d^{3}{\bar z}}\;
b\left(x^{+},{\bar x}-{\bar z}\right)
\psi_{+}^{\dagger}\left(x^{+},{\bar z}\right)
a\left(x^{+},{\bar z}-{\bar y}\right)
\Lambda_{-}.
\eeq
The contribution of the second order ${\cal B}_{2}$ is built out of two components, as the equation (\ref{expfmionr2}) indicates. Not complicated algebraic computations allow us to obtain both these expressions. By applying the equations (\ref{gtildematrixelement1}) we may put 
\beq
\label{2orderpart1}
\int_{R^{3}}\!{d^{3}{\bar z}_{1}}\!
\int_{R^{3}}\!{d^{3}{\bar z}_{2}}\!\;
{\cal F}_{0}^{-1}\!\left({\bar x}-{\bar z}_{1}\right)
{\cal{\tilde G}}_{2}\!\left({\bar z}_{1},{\bar z}_{2}\right)
{\cal F}_{0}^{-1}\!\left({\bar z}_{2}-{\bar y}\right)=
\eeq
\beqs
=\left(
\begin{array}{ccccc}
0&\;\;-H_{5}\left(x^{+},{\bar x},{\bar y}\right)&
\;\;0&\;\;0&\;\;0\\
H_{5}\left(x^{+},{\bar x},{\bar y}\right)&
\;\;0&\;\;0&\;\;0&\;\;0\\
0&\;\;0&\;\;0&\;\;0&\;\;0\\
0&\;\;0&\;\;0&\;\;0&\;\;0\\ 
0&\;\;0&\;\;0&\;\;0&\;\;0\\
\end{array}
\right).
\eeqs
Respectively, taking the pattern (\ref{gtildematrixelement2}), we have  
\beq
\label{2orderpart2}
\int_{R^{3}}\!{d^{3}{\bar z}_{1}}\dots\!
\int_{R^{3}}\!{d^{3}{\bar z}_{4}}\!\;
{\cal F}_{0}^{-1}\!\left({\bar x}-{\bar z}_{1}\right)
{\cal{\tilde G}}_{1}\!\left({\bar z}_{1},{\bar z}_{2}\right)
{\cal F}_{0}^{-1}\!\left({\bar z}_{2}-{\bar z}_{3}\right)
{\cal{\tilde G}}_{1}\!\left({\bar z}_{3},{\bar z}_{4}\right)
{\cal F}_{0}^{-1}\!\left({\bar z}_{4}-{\bar y}\right)=
\eeq
\beqs
=\left(
\begin{array}{ccccc}
H_{1}\left(x^{+},{\bar x},{\bar y}\right)&
\;\;H_{2}\left(x^{+},{\bar x},{\bar y}\right)&
\;\;0&\;\;0&\;\;0\\
H_{3}\left(x^{+},{\bar x},{\bar y}\right)&
\;\;H_{4}\left(x^{+},{\bar x},{\bar y}\right)&
\;\;0&\;\;0&\;\;0\\
0&\;\;0&\;\;0&\;\;0&\;\;0\\
0&\;\;0&\;\;0&\;\;0&\;\;0\\ 
0&\;\;0&\;\;0&\;\;0&\;\;0\\
\end{array}
\right).
\eeqs
Both these equations make up the complete second order contribution for the inverse ${\cal F}^{-1}(x^{+},{\bar x},{\bar y})$ matrix 
\beq
\label{invfinr2}
{\cal B}_{2}\!\left(x^{+},{\bar x},{\bar y}\right)=
g^{2}
\left(
\begin{array}{ccccc}
H_{1}\left(x^{+},{\bar x},{\bar y}\right)&
\;\;H_{2}\left(x^{+},{\bar x},{\bar y}\right)+
H_{5}\left(x^{+},{\bar x},{\bar y}\right)&
\;\;0&\;\;0&\;\;0\\
H_{3}\left(x^{+},{\bar x},{\bar y}\right)-
H_{5}\left(x^{+},{\bar x},{\bar y}\right)&
\;\;H_{4}\left(x^{+},{\bar x},{\bar y}\right)&
\;\;0&\;\;0&\;\;0\\
0&\;\;0&\;\;0&\;\;0&\;\;0\\
0&\;\;0&\;\;0&\;\;0&\;\;0\\ 
0&\;\;0&\;\;0&\;\;0&\;\;0\\
\end{array}
\right).
\eeq
The contribution ${\cal B}_{2}$ consists of several functions, which are similar to already introduced set of them $F_{1}(x^{+},{\bar x},{\bar y})$ - $F_{5}(x^{+},{\bar x},{\bar y})$, but now, they are bi-linear in fermionic or scalar fields:   
\beq
\label{invhfunct1}
H_{1}\left(x^{+},{\bar x},{\bar y}\right)=
\eeq
\beqs
=-\!\int_{R^{3}}\!{d^{3}{\bar z}_{1}}\!
\int_{R^{3}}\!{d^{3}{\bar z}_{2}}\;
a\left(x^{+},{\bar x}-{\bar z}_{1}\right)
{\psi}_{+}\left(x^{+},{\bar z}_{1}\right)
b\left(x^{+},{\bar z}_{1}-{\bar z}_{2}\right)
{\psi}_{+}\left(x^{+},{\bar z}_{2}\right)
a\left(x^{+},{\bar z}_{2}-{\bar y}\right)
\Lambda_{-},
\eeqs
\beq
\label{invhfunct2}
H_{2}\left(x^{+},{\bar x},{\bar y}\right)=
\eeq
\beqs
=\int_{R^{3}}\!{d^{3}{\bar z}_{1}}\!
\int_{R^{3}}\!{d^{3}{\bar z}_{2}}\;
a\left(x^{+},{\bar x}-{\bar z}_{1}\right)
{\psi}_{+}\left(x^{+},{\bar z}_{1}\right)
b\left(x^{+},{\bar z}_{1}-{\bar z}_{2}\right)
{\psi}_{+}^{\dagger}\left(x^{+},{\bar z}_{2}\right)
a\left(x^{+},{\bar z}_{2}-{\bar y}\right)
\Lambda_{-},
\eeqs 
\beq
\label{invhfunct3}
H_{3}\left(x^{+},{\bar x},{\bar y}\right)=
\eeq
\beqs
=\int_{R^{3}}\!{d^{3}{\bar z}_{1}}\!
\int_{R^{3}}\!{d^{3}{\bar z}_{2}}\;
a\left(x^{+},{\bar x}-{\bar z}_{1}\right)
{\psi}_{+}^{\dagger}\left(x^{+},{\bar z}_{1}\right)
b\left(x^{+},{\bar z}_{1}-{\bar z}_{2}\right)
{\psi}_{+}\left(x^{+},{\bar z}_{2}\right)
a\left(x^{+},{\bar z}_{2}-{\bar y}\right)
\Lambda_{-},
\eeqs 
\beq
\label{invhfunct4}
H_{4}\left(x^{+},{\bar x},{\bar y}\right)=
\eeq
\beqs
=-\!\int_{R^{3}}\!{d^{3}{\bar z}_{1}}\!
\int_{R^{3}}\!{d^{3}{\bar z}_{2}}\;
a\left(x^{+},{\bar x}-{\bar z}_{1}\right)
{\psi}_{+}^{\dagger}\left(x^{+},{\bar z}_{1}\right)
b\left(x^{+},{\bar z}_{1}-{\bar z}_{2}\right)
{\psi}_{+}^{\dagger}\left(x^{+},{\bar z}_{2}\right)
a\left(x^{+},{\bar z}_{2}-{\bar y}\right)
\Lambda_{-},
\eeqs
\beq
\label{invhfunct5}
H_{5}\left(x^{+},{\bar x},{\bar y}\right)=
\sqrt{2}i\!
\int_{R^{3}}\!{d^{3}{\bar z}}\;
a\left(x^{+},{\bar x}-{\bar z}\right)
{\phi}^{2}\left(x^{+},{\bar z}\right)
a\left(x^{+},{\bar z}-{\bar y}\right)
\Lambda_{-}.
\eeq

Not complicated algebra indicates, that the pure matrix products, which occur in the third and fourth order contributions are just trivial:
\beq
\label{thirdtrivial}
\left({\cal F}_{0}^{-1}{\cal{\tilde G}}_{1}\right)\!
\left({\cal F}_{0}^{-1}{\cal{\tilde G}}_{1}\right)\!
\left({\cal F}_{0}^{-1}{\cal{\tilde G}}_{1}\right)\!
{\cal F}_{0}^{-1}\!
=0,\;\;\;
\left({\cal F}_{0}^{-1}{\cal{\tilde G}}_{1}\right)\!
\left({\cal F}_{0}^{-1}{\cal{\tilde G}}_{2}\right)\!
{\cal F}_{0}^{-1}\!
=0,\;\;\; 
\left({\cal F}_{0}^{-1}{\cal{\tilde G}}_{2}\right)\!
\left({\cal F}_{0}^{-1}{\cal{\tilde G}}_{1}\right)\!
{\cal F}_{0}^{-1}\!
=0,
\eeq
\beq
\label{fourthtrivial}
\left({\cal F}_{0}^{-1}{\cal{\tilde G}}_{2}\right)
\left({\cal F}_{0}^{-1}{\cal{\tilde G}}_{2}\right)
{\cal F}_{0}^{-1}
=0. 
\eeq
Thus, the full contribution of the third order in the powers of coupling constant $g$ to the inverse matrix ${\cal F}^{-1}$ vanishes. We can also observe, by the inspection the general structure (\ref{finverseperturbativen}) of the $n$th order contributions to the ${\cal F}^{-1}$, that in each these terms, at least one of the combinations (\ref{thirdtrivial}) or (and) (\ref{fourthtrivial}) is always present, if $n\ge{4}$. It simply means, that all higher orders of contributions to (\ref{finverseperturbative}) are 
\beq
\label{bgt3zero}
{\cal B}_{n}\Big|_{n\ge{3}}=0.
\eeq
For this reason, at the end, we derived finite result for the inverse matrix ${\cal F}^{-1}(x^{+},{\bar x},{\bar y})$ in the Yukawa interacting model with one fermionic and one scalar field
\beq
\label{finvfin}
{\cal F}^{-1}\!\left(x^{+},{\bar x},{\bar y}\right)=
\eeq
\beqs
=\left(
\begin{array}{ccccc}
{\cal F}^{-1}_{{11}_{{\mbox{ }}_{{\mbox{ }}_{\mbox{ }}}}}\!\!\!\!\!\!\left(x^{+},{\bar x},{\bar y}\right)&
\;\;\;{\cal F}^{-1}_{12}\!\left(x^{+},{\bar x},{\bar y}\right)&\;\;\;
{\cal F}^{-1}_{13}\!\left(x^{+},{\bar x},{\bar y}\right)&
\;\;\;0&\;\;\;a\left(x^{+},{\bar x}-{\bar y}\right)I\\
{\cal F}^{-1}_{{21}_{{\mbox{ }}_{{\mbox{ }}_{\mbox{ }}}}}\!\!\!\!\!\!\left(x^{+},{\bar x},{\bar y}\right)&
\;\;\;{\cal F}^{-1}_{22}\!\left(x^{+},{\bar x},{\bar y}\right)&\;\;\;
{\cal F}^{-1}_{23}\!\left(x^{+},{\bar x},{\bar y}\right)&
\;\;\;a\left(x^{+},{\bar x}-{\bar y}\right)I&\;\;\;0\\
{\cal F}^{-1}_{{31}_{{\mbox{ }}_{{\mbox{ }}_{\mbox{ }}}}}\!\!\!\!\!\!\left(x^{+},{\bar x},{\bar y}\right)&
\;\;\;{\cal F}^{-1}_{32}\!\left(x^{+},{\bar x},{\bar y}\right)&\;\;\;
b\left(x^{+},{\bar x}-{\bar y}\right)I&\;\;\;0&\;\;\;0\\
0_{{ }_{{\mbox{ }}_{{\mbox{ }}_{\mbox{ }}}}}&\;\;a\left(x^{+},{\bar x}-{\bar y}\right)I&
\;\;\;0&\;\;\;0&\;\;\;0\\ 
a\left(x^{+},{\bar x}-{\bar y}\right)_{{ }_{{\mbox{ }}_{{\mbox{ }}_{\mbox{ }}}}}\!\!\!\!\!\!\!I&
\;\;\;0&\;\;\;0&\;\;\;0&\;\;\;0\\
\end{array}
\right).  
\eeqs
The elements inside this array are determined by the functions: $a(x^{+},{\bar x})$, $b(x^{+},{\bar x})$, $c(x^{+},{\bar x})$ from the free inverse matrix ${\cal F}_{0}^{-1}(x^{+},{\bar x},{\bar y})$. They also depend on already introduced expressions for the functions 
$F_{1}(x^{+},{\bar x},{\bar y})$ - $F_{5}(x^{+},{\bar x},{\bar y})$ and the $H_{1}(x^{+},{\bar x},{\bar y})$ - $H_{5}(x^{+},{\bar x},{\bar y})$. Finally: 
\beq
\label{finvfinelem1}
{\cal F}^{-1}_{11}\left(x^{+},{\bar x},{\bar y}\right)=
g^{2}H_{1}\left(x^{+},{\bar x},{\bar y}\right),
\eeq
\beq
\label{finvfinelem2} 
{\cal F}^{-1}_{12}\left(x^{+},{\bar x},{\bar y}\right)=
c\left(x^{+},{\bar x}-{\bar y}\right)I-
gF_{1}\left(x^{+},{\bar x},{\bar y}\right)+
g^{2}H_{2}\left(x^{+},{\bar x},{\bar y}\right)+
g^{2}H_{5}\left(x^{+},{\bar x},{\bar y}\right), 
\eeq
\beq
\label{finvfinelem3} 
{\cal F}^{-1}_{13}\left(x^{+},{\bar x},{\bar y}\right)=
-gF_{2}\left(x^{+},{\bar x},{\bar y}\right), 
\eeq
\beq
\label{finvfinelem4}
{\cal F}^{-1}_{21}\left(x^{+},{\bar x},{\bar y}\right)=
-c\left(x^{+},{\bar x}-{\bar y}\right)I+
gF_{1}\left(x^{+},{\bar x},{\bar y}\right)+
g^{2}H_{3}\left(x^{+},{\bar x},{\bar y}\right)-
g^{2}H_{5}\left(x^{+},{\bar x},{\bar y}\right),
\eeq
\beq
\label{finvfinelem56}
{\cal F}^{-1}_{22}\left(x^{+},{\bar x},{\bar y}\right)=
g^{2}H_{4}\left(x^{+},{\bar x},{\bar y}\right), \;\;\;\;\;\;  
{\cal F}^{-1}_{23}\left(x^{+},{\bar x},{\bar y}\right)=
-gF_{3}\left(x^{+},{\bar x},{\bar y}\right), 
\eeq 
\beq
\label{finvfinelem78} 
{\cal F}^{-1}_{31}\left(x^{+},{\bar x},{\bar y}\right)=
-gF_{4}\left(x^{+},{\bar x},{\bar y}\right), \;\;\;\;\;\; 
{\cal F}^{-1}_{32}\left(x^{+},{\bar x},{\bar y}\right)=
-gF_{5}\left(x^{+},{\bar x},{\bar y}\right). 
\eeq
Derived in this way matrix ${\cal F}^{-1}(x^{+},{\bar x},{\bar y})$ contains complete information about the interaction, available to extract from the formula (\ref{matrixseries}), however it is rather difficult to say at the first glance, does the pattern (\ref{finvfin}), together with the functions (\ref{finvfinelem1}) - (\ref{finvfinelem78}) provides the absolutely exact solution of our problem. Discussing it in more physical context, we meet  the question, whether presented here method is only perturbative or not. Moreover, finite or infinite character of obtained result may depend on fact, that we handle with solvable models or much more composite theories. This questions should be studied separately.\\ 

\section{Dirac Brackets for Interacting Fields}
\label{dbym}

Our issue (\ref{finvfin}) may be straightforwardly employed to derive the integral contribution in the pattern (\ref{dirbracketdef}), which arises from the Dirac-Bergmann procedure. It allows to establish the light-front Dirac brackets including the interaction for the Yukawa model. By taking into considerations the light-front canonical Poisson brackets for our model: (\ref{canquant1}), (\ref{canquant2}), (\ref{canquant3}), the equations of the constraints: (\ref{constraintsprimary}), (\ref{constraintssecondary4}), (\ref{constraintssecondary5}) and the formula (\ref{dirbracketdef}) we could write down the final result. We present this in the form with the bispinor fields $\Psi_{+}$ and $\Psi_{-}$. At the end, for the Yukawa model with coupled fermion and scalar field we have group of the unchanged Dirac brackets in comparison to the non-interacting case, collected in Chapter \ref{ifmf}. Accordingly:  
\beq
\label{dirbracketpsippsidpf}
\left\{
\Psi_{+}\left(x^{+},{\bar x}\right) \; , \; \Psi^{\dagger}_{+}\left(x^{+},{\bar y}\right)
\right\}_{D}=
\left\{
\Psi_{+}\left(x^{+},{\bar x}\right) \; , \; \Psi^{\dagger}_{+}\left(x^{+},{\bar y}\right)
\right\}_{P}=
{i\over{\sqrt{2}}}
\delta^{(3)}\left({\bar x}-{\bar y}\right)
\Lambda_{+},
\eeq
\beq
\label{dirbracketpsippsipf}
\left\{
\Psi_{+}\left(x^{+},{\bar x}\right) \; , \; \Psi_{+}\left(x^{+},{\bar y}\right)
\right\}_{D}=
\left\{
\Psi_{+}\left(x^{+},{\bar x}\right) \; , \; \Psi_{+}\left(x^{+},{\bar y}\right)
\right\}_{P}=0, 
\eeq
\beq
\label{dirbracketpsimpsipf}
\left\{
\Psi_{+}\left(x^{+},{\bar x}\right) \; , \; \Psi_{-}\left(x^{+},{\bar y}\right)
\right\}_{D}=
\left\{
\Psi_{+}\left(x^{+},{\bar x}\right) \; , \; \Psi_{-}\left(x^{+},{\bar y}\right)
\right\}_{P}=0, 
\eeq
\beq
\label{dirbracketphipsipf}
\left\{
\phi\left(x^{+},{\bar x}\right) \; , \; \Psi_{+}\left(x^{+},{\bar y}\right)
\right\}_{D}=
\left\{
\phi\left(x^{+},{\bar x}\right) \; , \; \Psi_{+}\left(x^{+},{\bar y}\right)
\right\}_{P}=0, 
\eeq
\beq
\label{dirbracketphiphif}
\left\{
\phi\left(x^{+},{\bar x}\right) \; , \; \phi\left(x^{+},{\bar y}\right)
\right\}_{D}=
{\underbrace{
\left\{
\phi\left(x^{+},{\bar x}\right) \; , \; \phi\left(x^{+},{\bar y}\right)
\right\}_{P}
}_{0}}+
b\left(x^{+},{\bar x}-{\bar y}\right),
\eeq
\beq
\label{dirbracketphiparmphif}
\left\{
\phi\left(x^{+},{\bar x}\right) \; , \; \partial^{y}_{-}\phi\left(x^{+},{\bar y}\right)
\right\}_{D}=
{\underbrace{
\left\{
\phi\left(x^{+},{\bar x}\right) \; , \; \partial^{y}_{-}\phi\left(x^{+},{\bar y}\right)
\right\}_{P}
}_{0}}+
\partial_{-}^{y}b\left(x^{+},{\bar x}-{\bar y}\right), 
\eeq
where the function $b(x^{+},{\bar x})$  and its derivative is determined by the equation (\ref{abdiffeq}). Of course, values of the Poisson brackets come from the canonical formalism. These brackets are entirely consistent with (anti-) commutators presented in Chapter \ref{sec1}. The second group of the Dirac brackets for the Yukawa model are determined by the details of the interaction between fermionic and scalar field, contained in the full matrix ${\cal F}^{-1}(x^{+},{\bar x},{\bar y})$ as the functions: (\ref{abdiffeq}), (\ref{cdiffeq}), (\ref{invfinr1elements1}) - (\ref{invfinr1elements5}) and (\ref{invhfunct1}) - (\ref{invhfunct5}). We have:  
\beq
\label{dirbracketpsimpsidpf}
\left\{
\Psi_{-}\left(x^{+},{\bar x}\right) \; , \; \Psi^{\dagger}_{+}\left(x^{+},{\bar y}\right)
\right\}_{D}=
{\underbrace{
\left\{
\Psi_{-}\left(x^{+},{\bar x}\right) \; , \; \Psi^{\dagger}_{+}\left(x^{+},{\bar y}\right)
\right\}_{P}
}_{0}}-
\eeq
\beqs
-{1\over{\sqrt{2}}}
\left\{
i\!\not\!\partial^{x}_{\perp}a\left(x^{+},{\bar x}-{\bar y}\right)-
ia\left(x^{+},{\bar x}-{\bar y}\right)
\left[M+g\phi\left(x^{+},{\bar y}\right)\right]
\right\}
\gamma^{+},
\eeqs
\beq
\label{dirbracketpsimpsidmf}
\left\{
\Psi_{-}\left(x^{+},{\bar x}\right) \; , \; \Psi^{\dagger}_{-}\left(x^{+},{\bar y}\right)
\right\}_{D}=
{\underbrace{
\left\{
\Psi_{-}\left(x^{+},{\bar x}\right) \; , \; \Psi^{\dagger}_{-}\left(x^{+},{\bar y}\right)
\right\}_{P}
}_{0}}+
\eeq
\beqs
+c\left(x^{+},{\bar x}-{\bar y}\right)\Lambda_{-}-
g{\tilde F}_{1}\left(x^{+},{\bar x},{\bar y}\right)+
g^{2}{\tilde H}_{2}\left(x^{+},{\bar x},{\bar y}\right)-
g^{2}{\tilde H}_{5}\left(x^{+},{\bar x},{\bar y}\right),
\eeqs
\beq
\label{dirbracketpsimpsimf}
\left\{
\Psi_{-}\left(x^{+},{\bar x}\right) \; , \; \Psi_{-}\left(x^{+},{\bar y}\right)
\right\}_{D}=
{\underbrace{
\left\{
\Psi_{-}\left(x^{+},{\bar x}\right) \; , \; \Psi_{-}\left(x^{+},{\bar y}\right)
\right\}_{P}
}_{0}}+
g^{2}{\tilde H}_{1}\left(x^{+},{\bar x},{\bar y}\right), 
\eeq
\beq
\label{dirbracketphipsimf}
\left\{
\phi\left(x^{+},{\bar x}\right) \; , \; \Psi_{-}\left(x^{+},{\bar y}\right)
\right\}_{D}=
{\underbrace{
\left\{
\phi\left(x^{+},{\bar x}\right) \; , \; \Psi_{-}\left(x^{+},{\bar y}\right)
\right\}_{P}
}_{0}}+
g{\tilde F}_{2}\left(x^{+},{\bar x},{\bar y}\right),
\eeq
wherein the functions with tildes are: 
\beq
\label{invfinr1elementstilda1}
{\tilde F}_{1}\left(x^{+},{\bar x},{\bar y}\right)=
2\sqrt{2}iM\!\!
\int_{R^{3}}\!{d^{3}{\bar z}}\;
a\left(x^{+},{\bar x}-{\bar z}\right)
{\phi}\left(x^{+},{\bar z}\right)
a\left(x^{+},{\bar z}-{\bar y}\right)
\Lambda_{-}=
F_{1}\left(x^{+},{\bar x},{\bar y}\right),
\eeq
\beq
\label{invfinr1elementstilda2}
{\tilde F}_{2}\left(x^{+},{\bar x},{\bar y}\right)=
-\!\int_{R^{3}}\!{d^{3}{\bar z}}\;
a\left(x^{+},{\bar x}-{\bar z}\right)
\left(\gamma^{+}\Psi_{+}\right)\left(x^{+},{\bar z}\right)
b\left(x^{+},{\bar z}-{\bar y}\right),
\eeq 
\beq
\label{invhfuncttilda1}
{\tilde H}_{1}\left(x^{+},{\bar x},{\bar y}\right)=
\eeq
\beqs
=-\!\int_{R^{3}}\!{d^{3}{\bar z}_{1}}\!
\int_{R^{3}}\!{d^{3}{\bar z}_{2}}\;
a\left(x^{+},{\bar x}-{\bar z}_{1}\right)
\left(\gamma^{+}{\Psi}_{+}\right)\left(x^{+},{\bar z}_{1}\right)
b\left(x^{+},{\bar z}_{1}-{\bar z}_{2}\right)
\left(\gamma^{+}{\Psi}_{+}\right)\left(x^{+},{\bar z}_{2}\right)
a\left(x^{+},{\bar z}_{2}-{\bar y}\right),
\eeqs
\beq
\label{invhfuncttilda2}
{\tilde H}_{2}\left(x^{+},{\bar x},{\bar y}\right)=
\eeq
\beqs
=\!\int_{R^{3}}\!\!{d^{3}{\bar z}_{1}}\!
\int_{R^{3}}\!{d^{3}{\bar z}_{2}}\;
a\left(x^{+},{\bar x}-{\bar z}_{1}\right)
\left(\gamma^{+}{\Psi}_{+}\right)\left(x^{+},{\bar z}_{1}\right)
b\left(x^{+},{\bar z}_{1}-{\bar z}_{2}\right)
\left({\Psi}_{+}^{\dagger}\gamma^{-}\right)\left(x^{+},{\bar z}_{2}\right)
a\left(x^{+},{\bar z}_{2}-{\bar y}\right),
\eeqs 
\beq
\label{invhfuncttilda5}
{\tilde H}_{5}\left(x^{+},{\bar x},{\bar y}\right)=
\sqrt{2}i\!
\int_{R^{3}}\!{d^{3}{\bar z}}\;
a\left(x^{+},{\bar x}-{\bar z}\right)
{\phi}^{2}\left(x^{+},{\bar z}\right)
a\left(x^{+},{\bar z}-{\bar y}\right)
\Lambda_{-}=
H_{5}\left(x^{+},{\bar x},{\bar y}\right).
\eeq
This group of the Dirac brackets has more folded structure than those from Chapters \ref{sec1} and \ref{ifmf}, due to fermionic - scalar coupling. They occur as the the next additional terms from matrix ${\cal F}^{-1}(x^{+},{\bar x},{\bar y})$, which include interaction. Of course we should remember, that this comparison is still accompanied by the final reservation from Chapter \ref{ifmf}, related to the anti-commutators  
$\{\Psi_{-}(x^{+},{\bar x}),\Psi^{\dagger}_{-}(x^{+},{\bar y})\}$ and $\{\Psi_{-}(x^{+},{\bar x}),\Psi_{-}(x^{+},{\bar y})\}$, descendant from relevant Dirac brackets.\\

\section{Conclusions}
\label{conc}

In this work we studied and applied the Dirac-Bergmann procedure with the aim to establish the Dirac brackets with interaction for the light-front Yukawa model in $D=1+3$ dimensions. This procedure of deriving the Dirac from Poisson brackets - consistent with the constraints attendant in the discussed model, sometimes shall cause the problems just on account the token interaction. We discussed from this point of view the light-front Yukawa model as a suitable scene for exploiting simple method leading to the solution of defined task. Finally, we included to the Dirac-Bergmann procedure the Yukawa interacting term between fermionic and scalar field. The main problem for such cases of the quantization has a mathematical character. The difficulties lie at this point, that we have to calculate the inverse matrix to the (\ref{fmatrixelements}) including the other matrices, differential operators and what is the most complicated, even the field operators for the enabled couplings. Proposed device comes down to the usage of the elementary matrix identity (\ref{matrixseries}). It has the form of the infinite, in principle, series and therefore, the obtained result seems to have the perturbative character, especially, that it contains the specific integral expressions. But the pure matrix structure of performed calculations causes, that the series is fast truncated for our model. It gives the contributions to the relevant Dirac brackets of the first and the second powers of the coupling constant $g$. Despite this slight uncertainty for the general, perturbative or not, character of presented solution, we finally calculated the Dirac brackets with interaction for the $D=1+3$ dimensional light-front Yukawa model.  The latter may be interesting from physical point of view for discussion on the structure and the properties of the (anti-) commutators of the interacting models on the light-front hyper-surface after the quantization. We compared obtained results to the computations inferred from the Heisenberg equations \cite{b15} and we underlined some delicate points of the light-front quantization. The open problem is, whether the matrix ${\cal F}^{-1}$ gives the complete and exact, whether only the approximate (perturbative) solution of our problem. However, the finite or infinite character of obtained result may strongly depend on fact, what kind of the theory or model we handle - solvable or not, for instance. This is a good area for further discussion.\\


\appendix{{\bf{Appendix: Light-Front Coordinates, Dirac Matrices and Field Operators}}}\\
\label{app1}

In this work we employed the light-front coordinates in $D=1+3$ dimensions: 
\beq
\label{lfcoord}
x=
\left(x^{+},x^{-},{\bf x}_{\perp}\right), \;\;\;\;\;\; 
x^{\pm}=
{1\over{\sqrt{2}}}\left(x^{0}\pm{x^{3}}\right), \;\;\;\;\;\;  
{\bf x}_{\perp}=
\left(x^{1},x^{2}\right). 
\eeq
The light-front and the anti-light-front hyper-surfaces $x^{+}=0$, respectively $x^{-}=0$, are described by the reduced coordinates, written as: 
\beq
\label{lfhypcoord}
{\bar x}=
\left(x^{-},{\bf x}_{\perp}\right)\simeq
\left(x^{+}=0,x^{-},{\bf x}_{\perp}\right), \;\;\;\;\;\;  
{\underline x}=
\left(x^{+},{\bf x}_{\perp}\right)\simeq
\left(x^{+},x^{-}=0,{\bf x}_{\perp}\right).
\eeq
According to the definitions (\ref{lfcoord}), the light-front metric tensor has following components: 
\beq
\label{gtensorcomp}
g_{+-}=1, \;\;\; \;\;\;
g_{++}=0=g_{--}, \;\;\;\;\;\; 
g_{{\pm}j}=0, \;\;\;\;\;\; 
g_{jk}=-\delta_{jk},\;\;\;\;\;\; 
j,k=1,2. 
\eeq
Therefore, the rules for lifting and and pulling down the index of the light-front four-vector coordinates are: 
\beq
\label{lifting}
a_{\pm}=a^{\mp}, \;\;\;\;\;\; 
a_{j}=-a^{j}, \;\;\;\;\;\; 
j=1,2. 
\eeq
In the case of the light-front metric tensor, the scalar product of the four-vectors obeys: 
\beq
\label{scalprod}
a\cdot{b}=
a_{+}b_{-}+
a_{-}b_{+}-{\bf a}_{\perp}\cdot{\bf b}_{\perp}, \;\;\;\;\;\;  
{\bf a}_{\perp}\cdot{\bf b}_{\perp}=
a_{j}b_{j}, \;\;\;\;\;\; 
j=1,2,  
\eeq
wherein the Einstein notation is used. In this regard, the light-front Dirac slash satisfies:   
\beq
\label{slash}
\not\!a=
a_{\mu}\gamma^{\mu}=
a_{+}\gamma^{+}+a_{-}\gamma^{-}-
\not\!{\bf a}_{\perp}, \;\;\;\;\;\; 
\not\!{\bf a}_{\perp}=
a_{j}\gamma^{j}, \;\;\;\;\;\;
\mu=+,-,j, \;\;\;\;\;\; 
j=1,2.
\eeq
The light-front derivatives: 
\beq
\label{derivlf}
\partial_{\pm}=
{{\partial}\over{\partial{x^{\pm}}}}, \;\;\;\;\;\; 
\partial_{j}=
{{\partial}\over{\partial{x^{j}}}}, \;\;\;\;\;\; 
j=1,2,  
\eeq
fix the Laplace and the d'Alembert operators: 
\beq
\label{dal}
\Delta_{\perp}=
\partial_{j}\partial_{j}, \;\;\;\;\;\;   
\Box=
2\partial_{+}\partial_{-}-\Delta_{\perp}, \;\;\;\;\;\; 
j=1,2. 
\eeq
We have also entirely analogous to the (\ref{lfcoord}) definitions for the gamma matrices: 
\beq
\label{defgammapm}
\gamma=
\left(\gamma^{+},\gamma^{-},{\bf\gamma}_{\perp}\right), \;\;\;\;\;\; 
\gamma^{\pm}=
{1\over{\sqrt{2}}}\left(\gamma^{0}\pm{\gamma^{3}}\right), \;\;\;\;\;\;  
{\bf\gamma}_{\perp}=\left(\gamma^{1},\gamma^{2}\right). 
\eeq
These arrays obey following properties: 
\beq
\label{gammamatrixprop}
\left(\gamma^{\pm}\right)^{2}=0, \;\;\;\;\;\;  
\left(\gamma^{\pm}\right)^{\dagger}=\gamma^{\mp}, \;\;\;\;\;\; 
\left(\gamma^{j}\right)^{\dagger}=-\gamma^{j}, \;\;\;\;\;\; 
j=1,2. 
\eeq
In this work we applied decomposition of the $\Psi$, the fermionic bispinor 
\beq
\label{spindecomppsi}
\Psi=
\Psi_{+}+\Psi_{-},   
\eeq
onto two pure spinors $\psi_{+}$ and $\psi_{-}$ with up ($+$) and down ($-$) indexes, respectively:
\beq
\label{spinors}
\Psi=
\left(
\begin{array}{c}
\psi_{+}\\
\psi_{-}\\
\end{array}
\right), \;\;\;\;\;\; 
\Psi_{+}=
\left(
\begin{array}{c}
\psi_{+}\\
0\\
\end{array}
\right)\simeq\psi_{+}, \;\;\;\;\;\; 
\Psi_{-}=
\left(
\begin{array}{c}
0\\
\psi_{-}\\
\end{array}
\right)\simeq\psi_{-}. 
\eeq
For this goal we introduced two constant bispinors: 
\beq
\label{uconstpmspin}
u_{+}=
\left(
\begin{array}{c}
1\\
0\\
\end{array}
\right), \;\;\;\;\;\;
u_{-}=
\left( 
\begin{array}{c}
0\\
1\\
\end{array}
\right), \;\;\;\;\;\; 
\psi_{\pm}=u_{\pm}^{\dagger}\Psi,
\eeq
which made the definition of the matrices $\Lambda_{\pm}$, as they were implemented as following tensor products: 
\beq
\label{tenprodlambdapm}  
u_{\pm}u^{\dagger}_{\pm}=
\Lambda_{\pm}=
u^{\dagger}_{\pm}u_{\pm},
\eeq
displaying explicitly the form: 
\beq
\label{expliclambda}
\Lambda_{+}=
\left(
\begin{array}{cc} 
1&\;\;0\\
0&\;\;0\\
\end{array}
\right), \;\;\;\;\;\; 
\Lambda_{-}=
\left(
\begin{array}{cc} 
0&\;\;0\\
0&\;\;1\\
\end{array}
\right) 
\eeq
and finally permitting us to put: 
\beq
\label{psipmaslambdapsi}
\Psi_{+}=\Lambda_{+}\Psi, \;\;\;\;\;\;  
\Psi_{-}=\Lambda_{-}\Psi.
\eeq
The arrays $\Lambda_{\pm}$ are projective: 
\beq
\label{projprop}
\Lambda_{\pm}^{2}=
\Lambda_{\pm}, \;\;\;\;\;\; 
\Lambda_{\pm}\Lambda_{\mp}=0=
\Lambda_{\mp}\Lambda_{\pm}, \;\;\;\;\;\; 
\Lambda_{+}+\Lambda_{-}=I 
\eeq 
and Hermitian:
\beq
\label{hermit}
\Lambda_{\pm}^{\dagger}=
\Lambda_{\pm}.
\eeq
It is comfortable to express the $\Lambda_{\pm}$ matrices by the $\gamma^{\pm}$ arrays   
\beq
\label{lambdabygamma}
\Lambda_{\pm}=
{1\over 2}
\gamma^{\mp}\gamma^{\pm}.
 \eeq
These $\gamma^{\pm}$ may be defined in the analogous way to the $\Lambda_{\pm}$ matrices    
\beq
\label{gammabyu}
u_{\mp}u^{\dagger}_{\pm}=
\gamma^{\pm}=
u^{\dagger}_{\pm}u_{\mp}
\eeq
and therefore, their explicit patterns satisfy 
\beq
\label{gammmatabl}
\gamma^{+}=
\left(
\begin{array}{cc} 
0&\;\;0\\
1&\;\;0\\
\end{array}
\right), \;\;\;\;\;\; 
\gamma^{-}=
\left(
\begin{array}{cc} 
0&\;\;1\\
0&\;\;0\\
\end{array}
\right). 
\eeq
Accordingly, the matrices $\gamma^{\pm}$ commute with the arrays $\Lambda_{\pm}$ and obey 
\beq
\label{gammalambdainterm}
\gamma^{\pm}\Lambda_{\pm}=
\gamma^{\pm}=
\Lambda_{\mp}\gamma^{\pm}.
\eeq
The $\gamma^{\pm}$ arrays act on the bispinor $\Psi$ and give: 
\beq
\label{gammapsi}
\gamma^{+}\Psi=
\left(
\begin{array}{c}
0\\
\psi_{+}\\
\end{array}
\right), \;\;\;\;\;\;
\gamma^{-}\Psi=
\left(
\begin{array}{c}
\psi_{-}\\
0\\
\end{array}
\right), \;\;\;\;\;\;
\gamma^{\pm}\Psi_{\mp}=0.
\eeq
All the above formulas played very useful role for the intermediate calculations in this work.\\

\end{document}